\documentstyle[12pt]{article}

\begin{document}
\begin{flushright}
{BI-TP 93/38}
\end{flushright}
\begin{flushright}
{ETH-TH/93-25}
\end{flushright}
\begin{flushright}
{July 1993}
\end{flushright}
\begin{center}
{\bf RESUMMING THE EFFECTIVE ACTION}\footnote{Presented by A.L. at
"Quark Matter 93", B$\ddot{o}$rlange, Sweden}
\end{center}
\begin{center}
{\bf Andrei Leonidov\footnote{Alexander von Humboldt Fellow}$^{(a)}$
and Andrei Zelnikov$^{(b)}$}
\end{center}
\begin{center}
{\it (a) Physics Department\\ University of Bielefeld D-4800,
Bielefeld,Germany\\ and \\Theoretical Physics Department\\ P.N.Lebedev
Physics Institute, 117924 Leninsky pr.53, Moscow, Russia}\\
\medskip

{\it (b) Institut fur Theoretische Physik\\ ETH-Honggerberg
CH-8093, Zurich, Switzerland\\and\\P.N.Lebedev Physics Institute\\ 117924
Leninsky
pr.53, Moscow,
Russia}
\end{center}
\begin{abstract}
At the simple example of a massless scalar field propagating in the
static background we study the resummed expressions for the effective
action at zero and finite temperature that are free from a usual
infrared sickness of the effective action induced by massless particles.
\end{abstract}

\newpage

\begin{center}
1. \it{Introduction}
\end{center}

In this note we discuss at simple examples the ways of dealing with the
well-known problem of the infrared sickness of the one-loop effective
action for the massless test particles propagating in the external
field.

Let us first remind the essence of the above-mentioned problem and
consider the simplest case of a scalar field $\varphi(x)$, propagating in the
background
field $\Phi (x)$. The corresponding Lagrangian has the
form
\begin{equation}
L[\varphi,\Phi]={1 \over 2}\varphi(x)(-\Box+m^2+\Phi^{2} (x))\varphi(x)
\end{equation}
where $\Box$ is a free Laplacian and the corresponding effective action for
$\Phi(x)$
induced by
the test particles $\varphi(x)$ (or using the terminology from
statistical mechanics, the induced entropy of the external field) at the
one loop level is
\begin{equation}
W[\Phi]=-{1 \over 2}\int_{0}^{\infty}{ds \over s} e^{-m^2 s}Tr(e^{-(-\Box+
\Phi^2 (x))s}-e^{\Box s})
\end{equation}
where we have used the proper time representation and the interesting
contribution is
isolated by substracting the free
propagator  contribution. Certainly the above expression should be
regularized. In the following we shall exploit the following variant of the
$\zeta$-regularization:
\begin{equation}
LogDetA=-{d \over \delta}(({M^{2\delta} \over \Gamma (\delta)})
\int_{0}^{\infty} {ds \over s^{1-\delta}}Tre^{-As})_{\delta=0}
\end{equation}
where $M$ is an ultraviolet cutoff. In a situation when the external field
$\Phi (x)$ is essentially inhomogeneous one is usually confined to using
only a certain number of terms in the expansion of the trace of the heat kernel
\begin{equation}
TrK(s)=Tr e^{-(-\Box+\Phi^{2} (x))s}
\end{equation}
in the powers of the proper time $s$
\begin{equation}
TrK(s)=\sum_{n=0}^{\infty}b_{-\omega+n}(\Phi)s^{-\omega+n}
\end{equation}
 where $2\omega$ is a (eucledian) space-time dimension. We see that in the
massless case $m=0$ starting from some term in the heat kernel expansion
the corresponding effective action is term by term divergent at the
upper limit of the integration over $s$ and therefore the effective action
is an operationally ill-defined quantity in the infrared domain.
 The same problem arises in the finite temperature calculations of the
free energy of a system of massless particles in the inhomogeneous
static external field. Namely, we have for the free energy of this system at
the one-loop level
\begin{equation}
-\beta F={1 \over 2}\int_{0}^{\infty}{ds \over s}{ \beta \over (4\pi s)^
{1 \over 2}}\sum_{-\infty}^{\infty} e^{-{n^2 \beta^{2} \over 4s}} Tr K
_{2\omega-1}(s)
\end{equation}
which after using the Poisson resummation formula
\begin{equation}
\sum_{n=-\infty}^{\infty}e^{-{n^2 \beta^{2} \over 4s}}=
({4\pi s \over \beta^{2}})^{{1 \over 2}}
\sum_{n=-\infty}^{\infty}e^{-{4\pi^{2} s \over \beta^{2}} n^2}
\end{equation}
takes the form
\begin{equation}
-\beta F={1 \over 2}\int_{0}^{\infty}{ds \over s}(1+2\sum_{n=1}^{\infty}
 e^{-{4\pi^{2} s \over \beta^{2}} n^2}) Tr K_{2\omega-1}(s)
\end{equation}
where $\beta$ is an inverse temperature
(let us recall that here the eucledian time is
compactified on a circle of a length $\beta$). We see that for $n=0$
(entropy) term when using a usual semiclassical expansinon for the spatial heat
kernel $K_{2\omega-1}(s)$ we face the same problem  of a term by
term infrared divergence at the upper limit of the integration over $s$.

This unpleasant situation (which is of course arising for any type of the
massless test particle: gluons, gravitons, etc.) forces us to find a way
to give an operational definition for the effective action by using such
approximations for the heat kernel $K(s)$ that the
corresponidng effective action would be finite. In the following we
shall consider two such approximations actually corresponding to the two
different types of resumming certain infinite sequencies of terms in
the heat kernel.

The first method corresponds to a summation of some sequence of terms of
all orders in the external field $\Phi (x)$ using the analogy between
the trace of the heat kernel and the partition function, for which the
corresponidng resummation has been discussed in the literature [1]. We find
that for the simple external scalar field configurations the infrared
problem is cured and it is possible to get an infrared-finite answer for
the effective action (entropy) already for the simplest of previosly
discussed resummation for the partition function.

The second  method corresponds to summing all derivatives for the terms
having some definite power in the external field. This corresponds to a
calculation of a nonlocal effective action [2-5]. In this method it was proved
both at zero [2-4] and finite temperature [5] that this procedure allows to get
an
infrared-finite answer for the effective action (free energy). Below  we
illustrate this situation - again using the simplest localized scalar
field configurations. Here a (small)  new step is using the explicitely
nonlocal formulas for the effective action .

Let us notice, that formally it is possible to exponentiate the nonlocal
terms in the effective action too (at least for scalar [6] and electromagnetic
[7]
interactions). However it is not at all clear whether this method could
be used for actual computations. We hope to return to the analysis of
this question elsewhere.

 Let us also mention that in estimating the effective action one can use
the method of term-by-term infrared regularization of the effective
action using the series (5) and optimizing with respect to a cutoff [8],
but in this case it seems to be difficult to trace the interrelations
between the relevant scales to motivate the expansion parameter.

\begin{center}
2. \it{Resumming the potential}
\end{center}

The origin of the above-described infrared catastrophy is clearly in
using the expansion of the trace of the heat kernel in the powers of the
proper time. Thus an obvious idea is to use an expression for the

trace of the heat kernel which is (at leat approximately) exponential in
the external field. In fact the expressions of such type have already
been discussed in the statistical physics context [1], where the analogue
of the trace of the heat kernel is a partition function, the inverse
temperature
being the analogue of the proper time. As our main purpose here is to
illustrate the corresponding possibilities at the simple solvable
examples, we shall use just the original version of the formula for the
partition function as calculated by Goldberger and Adams [1]. In our
notation the expression for the trace of the heat kernel for the
massless scalar particle propagating in the external field $\Phi (x)$ reads
\begin{equation}
Tr(K(s)-K_{0}(s))={1 \over (4\pi s)^{\omega}}\int d^{2\omega} x
(e^{-\Phi^{2}(x)s}(1-s^2 ({1 \over 6}\Box \Phi^2 (x)
-{1 \over 12} (\nabla \Phi^2 (x))^2+...)-1)
\end{equation}
In the finite temperature case (as seen from Eq.(8)) this trace should be
taken over $2\omega-1$-dimensional space (for the static external field
configurations).

Let us now consider a localized spherically symmetric static external field
configuration in $3$ dimensions
\begin{equation}
\Phi (r) = {\Phi_{0} \over (1+r/a)^{2}}
\end{equation}
and first calculate the effective action in the zero temperature
case. Substituting the external field configuration (10) into the
expression for the heat kernel Eq.(9)
we get in the leading approximation for the heat kernel ($\omega=2$):
\begin{equation}
Tr(K(s)-K_{0}(s))={\pi \tau \Phi_{0}^{4} a^3 \over (4\pi \sigma)^{2}}
(\sigma^{{1 \over 4}} \gamma (-{1 \over 4},\sigma)
-2\sigma^{{1 \over 2}} \gamma(-{1 \over 2}, \sigma)
+\sigma^{{3 \over 4}} \gamma (-{3 \over 4},\sigma)+{4 \over 3})
\end{equation}
where $\tau$ is an eucledian time, $\gamma (\alpha,\sigma)$ is an
incomplete gamma-function and we have introduced the dimensionless
variable $\sigma=\Phi_{0}^{2} s$. The terms originating from the
derivative corrections in Eq.(9) can also be calculated, and finally one
gets for the effective action
\begin{equation}
W[\Phi_{0},a,M]=-{1 \over 64\pi^2}\tau ({4 \over 3}\pi a^3)\Phi_{0}^4
{1 \over 35}Log{M^2 \over \Phi_{0}^2} (1+{70 \over 63} {1 \over
\Phi_{0}^2 a^2}) + ...
\end{equation}
We can conclude that in the simplest approximation for the exponentiated
heat kernel the basic difference from the familiar constant external
field case (the first term in  the above expresion would just correspond to a
usual
effective potential) is the appearence of the effective volume $a^3$ of
the localized configuration instead of the total spatial volume
$V^{(3)}$. The logarithmic factor still depends only on the external
field amplitude  and taking into account the terms with derivatives
results in the expansion in the inverse powers of $\Phi_{0}^2 a^2$.

Let us now calculate the
free energy of a massless scalar field propagating in the static
backgorund field configuration Eq.10. Combining the Eqs. 8 and 11 we
get:
\begin{equation}
-({F \over T})={(\Phi_{0} a)^{3} \over 90}\\
-{2 \over 3} (\Phi_{0} a)^3 \sum_{n=1}^{\infty}(1+\mu n^2)^{3
\over 2}f(\mu,n)
\end{equation}
where $\mu={4\pi^2T^2 \over \Phi_{0}^2}$ and
\begin{eqnarray}
f(\mu,n^2) &=& ~_{2}F_{1}(1;-{3 \over 2};{3 \over 4};{1 \over 1+\mu n^2})
-~_{2}F_{1}(1;-{3 \over 2};{1 \over 2};{1 \over 1+\mu n^2}) \nonumber\\
&& +{1 \over 3}~_{2}F_{1}(1;-{3 \over 2};{1 \over 4};{1 \over 1+\mu
n^2}) - {1 \over 3} ({\mu n^2 \over 1+\mu n^2})^{3 \over 2}
\end {eqnarray}
where $~_{2}F_{1}(a;b;c;x)$ is a hypergeometric function and for
simplicity
we took into account only the leading exponential term (the
derivative corrections can be derived in a completely analogeous way).
  From this expression one can work out the low- and high- temperature
expanisons in a standard way.

\begin{center}
2. \it{Summing the derivatives}
\end{center}

Let us now turn to the second possibility of constructing an
infrared-safe approximation for the calculation of the effective action
for massless test particles. This method corresponds to a summation of
all derivative terms for a given power of the external field. The
resulting effective action is therefore an essentially nonlocal object.
In a pioneering paper [3] Barvinsky and Vilkovisky have shown that this
procedure provides infrared-convergent integrals for the effective
action in all orders in the external field. Later this technique was
generalized to a finite-temperature case [5]. In [2-4,5] all the formfactors
were written as infinite series in the free laplacian. This is quite
suitable for the formal purposes, but for physical applications it is
desirable to have the expressions which are explicitely nonlocal in the
external fields, thus facilitating the analysis of the impact of the
scales (correlation lengths) set up by the external field. Below we
shall analyse the explicitely nonlocal effective action for the same
case of a massless scalar field propagating in an localized external
field configuration.

For the trace of the heat kernel we have a general expansion
\begin{equation}
TrK(s)=\sum_{n=0}^{\infty} TrK_{n}(s)
\end{equation}
where
\begin{equation}
TrK_{n}(s)={s^{n} \over n} \int_{\alpha_{i} \geq 0} d^{n}\alpha
\delta (1-\sum_{i=1}^{n} \alpha_{i})
Tr[Ve^{s\alpha_{1}\Box}...Ve^{s\alpha_{n}\Box}]
\end{equation}
and for the external field potental $V$ we shall take an
$O(3)$-symmetric external field configuration
\begin{equation}
V=\Phi_{0}^2 e^{-{r^2 \over 2a^2}}
\end{equation}
where the choice of the configurattion to consider is dictated by
a computational simplicity. For the effective action at zero temperature
we get in the third order in the external field
perturbation:
\begin{equation}
W={1 \over 64\pi^{{1 \over 2}}} {\tau \over a}(\Phi_{0}^2 a^2)^2 Log({1
\over M^2 a^2}) - {c \over 192 (2\pi)^{1 \over 2}}({\tau \over a} (\Phi_{0}^2
a^2)^3)
\end{equation}
where the constant $c=1.30348$ was obtained by numerical integration. We see
that the
basic difference of this answer from that obtained in
the first section is that the charge renormalization logarithm is now
saturated by the slope of the field, and not by its amplitude. For the
free energy one gets in the same limit
\begin{eqnarray}
-\beta F & = & {\pi \over 16}(\Phi_{0} a)^4 \nonumber\\
& & +{\pi^{{5 \over 2}} \over 4}
(\Phi_{0} a)^4 (aT)^2
\sum_{n=1}^{\infty} n^2 \int_{0}^{1}d\alpha\alpha^{-{3 \over 2}}
(1-\alpha)^{-{3 \over 2}}\Psi({3 \over 2},2,{4\pi^2a^2T^2n^2 \over
 \alpha(1-\alpha)})
\end{eqnarray}
where $\Psi(a,c;x)$ is a confluent hypergeometric function. This
expression can serve as a starting point for constructing the low- and
high- temperature expansions by standard methods.

\begin{center}
4. \it{Aknowledgments}
\end{center}
A.L is grateful to Prof. H. Satz for kind hospitality at the University
of Bielefeld where this paper was finished.
His work was
partially supported by the Russian Fund for Fundamental Research, Grant
93-02-3815.
A.Z. is grateful to Prof. C. Schmid for kind hospitality at
the Institut for Theoretical Physics, ETH-Honggerberg (Zurich).
His work was partially supported by Soros Fund
and American Astronomical Society.

\newpage
\begin{center}
{\it References}
\end{center}

1. M.L.Goldberger, E.N.Adams. {\it{Journ. Chem. Phys.}} {\bf{20}} (1952),
240;

2. A.O.Barvinsky, G.A.Vilkovisky {\it{Nucl.Phys}} {\bf{B282}} (1987),163;

3. A.O.Barvinsky, G.A.Vilkovisky {\it{Nucl. Phys}} {\bf{B333}} (1990),471;

4. A.O.Barvinsky, G.A.Vilkovisky {\it{Nucl.Phys}} {\bf{B333}} (1990),512;

5  A.V.Leonidov, A.I.Zelnikov {\it{Phys.Lett}} {\bf{B276}} (1992),122;

6. Y.Fujiwara, T.A.Osborn and S.F.J. Wilk {\it{Phys.Rev.}} {\bf{A25}}
(1982),14;

7. A.O.Barvinsky, T.A.Osborn Manitoba University preprint {\it MANI-92-01,
May 1992.}

8. D.I.Diakonov, V.Yu.Petrov and A.V.Yung {\it{Phys.Lett.}}{\bf{B130}}
(1983),385; {\it{Sov.J.Nucl.Phys.}}{\bf{39}} (1984), 150.

\end{document}